\documentclass[10pt, conference, letterpaper]{IEEEtran}
\IEEEoverridecommandlockouts
\usepackage{cite}
\usepackage{amsmath,amssymb,amsfonts}
\usepackage{algorithm}
\usepackage{algorithmicx}
\usepackage{graphicx}
\usepackage{textcomp}
\usepackage{xcolor}
\usepackage{subfigure}
\usepackage{lipsum}
\usepackage{algpseudocode}
\usepackage{tikz}
\newcommand*\circled[1]{\tikz[baseline=(char.base)]{
            \node[shape=circle,draw,inner sep=1pt] (char) {#1};}}
\usepackage{enumitem}

\usepackage{balance} 

\newcommand{\model}{TPAoI~}

\def\BibTeX{{\rm B\kern-.05em{\sc i\kern-.025em b}\kern-.08em
    T\kern-.1667em\lower.7ex\hbox{E}\kern-.125emX}}
\begin{document}


\title{TPAoI: Ensuring Fresh Service Status at the Network Edge in Compute-First Networking
}



\author{
	\IEEEauthorblockN{
		Haosheng He 
		, Jianpeng Qi 
		, Chao Liu 
		, Junyu Dong 
		and Yanwei Yu$^{\ast}$
        } 
	\IEEEauthorblockA{School of computer science and technology\\ Ocean University of China, Qingdao, China\\ Email: hhs@stu.ouc.edu.cn, \{qijianpeng, liuchao, dongjunyu, yuyanwei\}@ouc.edu.cn}
    \thanks{*Corresponding authors}
}

\maketitle

\begin{abstract}
In compute-first networking, maintaining fresh and accurate status information at the network edge is crucial for effective access to remote services. This process typically involves three phases: Status updating, user accessing, and user requesting. However, current studies on status effectiveness, such as  Age of Information at Query (QAoI), do not comprehensively cover all these phases. Therefore, this paper introduces a novel metric, TPAoI, aimed at optimizing update decisions by measuring the freshness of service status. The stochastic nature of edge environments, characterized by unpredictable communication delays in updating, requesting, and user access times, poses a significant challenge when modeling. To address this, we model the problem as a Markov Decision Process (MDP) and employ a Dueling Double Deep Q-Network (D3QN) algorithm for optimization. Extensive experiments demonstrate that the proposed TPAoI metric effectively minimizes AoI, ensuring timely and reliable service updates in dynamic edge environments. Results indicate that TPAoI reduces AoI by an average of 47\% compared to QAoI metrics and decreases update frequency by an average of 48\% relative to conventional AoI metrics, showing significant improvement.
\end{abstract}


\section{Introduction}\label{sect-intro}
A critical challenge in compute-first networking is ensuring the freshness and accuracy of monitored data, which directly impacts the effectiveness of accessing remote services \cite{jianpeng2024, cfn2019icn}. Central to addressing this challenge is the Age of Information (AoI), a pivotal metric that quantifies the timeliness of information \cite{yates2021}. AoI measures the elapsed time from data generation to its retrieval, playing a crucial role in assessing system performance. This metric is particularly vital in contexts requiring rapid response capabilities, such as Internet of Things (IoT) \cite{abbas2023comprehensive}, healthcare systems \cite{ling2022age}, and autonomous driving \cite{xu2022aoi}.
\begin{figure}[htbp]
\centerline{}
\includegraphics[width=0.50\textwidth]{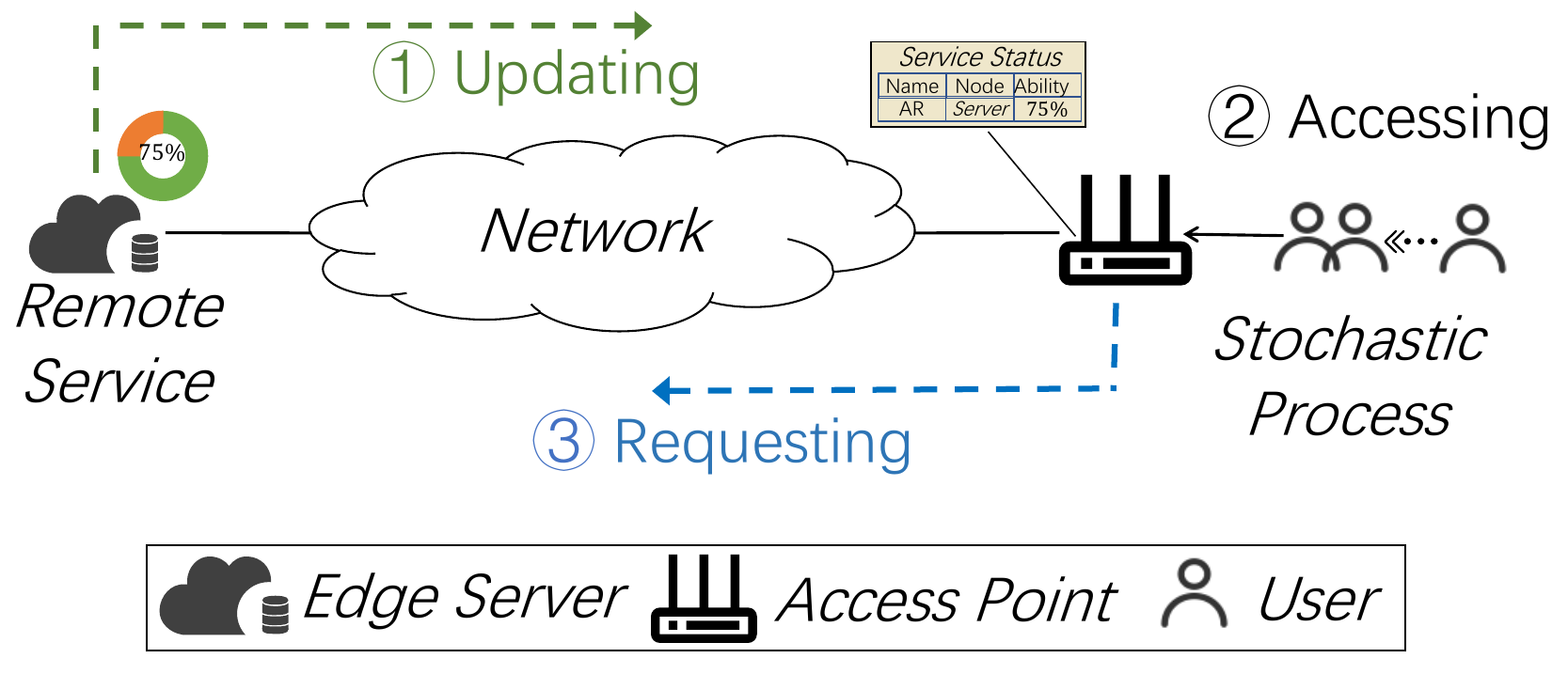}
\caption{Illustration of compute-first networking}
\label{system_model}
\end{figure}

\figurename~\ref{system_model} provides a general process of service provisioning in the compute-first networking domain. In this process, users at the network edge can access the updated status stored on access points (APs) and then request the remote service. It can be broadly divided into three phases \cite{jianpeng2024, cpnsurvey2024}: \circled{1} \textit{Updating (or forwarding)}, an edge server distributes service status updates (e.g., its recently supported concurrent capacity) to the APs via the network. \circled{2} \textit{Accessing}, users, upon accessing the APs, can evaluate the updated status to determine whether to send a request to the remote service. During this phase, the process can be modeled as a stochastic process. \circled{3} \textit{Requesting (or backward requesting)}, AP sends the request to remote service, incurring a random communication delay. 

As commonly pointed out in AoI analysis, a lower status updating frequency may cause service monitoring errors, while a higher frequency can lead to communication burdens and energy budget \cite{yates2021}. When ensuring the status updating frequency, as illustrated, there are three stochastic phases involved—\textit{updating, accessing, and requesting}. During the updating and requesting phases, bandwidth could be changing, affecting the transmitting delays. Meanwhile, in the accessing phase, the user accessing times are stochastic and not easy to model. Therefore, given those dynamic edge environments \cite{easiei2024iotj}, ensuring the effectiveness of the service status considering those three phases could be challenging.

Previous work has somewhat ignored this integrity. In early AoI studies, most literature assumes that users continuously access the APs and consider the optimal AoI at every moment \cite{yates2021, qaoi2022tcomm}. However, in real-world scenarios, user access behaviors are stochastic. Some studies, such as QAoI \cite{qaoi2022tcomm} and EAoI \cite{eaoi2019infocom}, focus on the AoI at user access times, updating just before user access, which significantly reduces resource costs. Nevertheless, they assume that the status sent by the edge server can quickly reach the access point or its buffer. In practical applications, transmission delay (including the updating and requesting) can significantly impact system performance, especially under heavy network loads or long transmission distances. Ignoring the effects of transmission delay fluctuations may result in status updates arriving after the user access.

To fill that gap, we propose a new age metric called \underline{T}hree-\underline{P}hase AoI (TPAoI), which incorporates phases such as updating, requesting, and user stochastic accessing. Given the dynamic nature of edge environments, precisely depicting them is complex, therefore we model the problem of minimizing the \model as an MDP and optimize it using a model-free approach in deep reinforcement learning (DRL) methods. Specifically, we first design five sub-states according to the three phases to represent the environments and then implement an algorithm based on D3QN to accurately capture the dynamic behaviors. Results show that our solution can reduce both the AoI and updating frequency effectively, reducing AoI by an average of 47\% compared to QAoI, and decreasing the update frequency by an average of 48\% over conventional AoI.

The main contributions in this work are as follows:

\begin{itemize}
    \item We comprehensively consider the AoI during three phases, including status updating, user accessing, and user requesting, and propose a metric, TPAoI, to measure the freshness of the service status.
    \item To handle the dynamic edge environments, we formulate the system model as an MDP and carefully design the state to satisfy the MDP property. Subsequently, we employed the Dueling Double DQN (D3QN) algorithm to optimize the MDP, thereby effectively capturing the environmental dynamics and showcasing the method’s robustness.
    \item We conducted extensive experiments to validate the effectiveness of the proposed method in solving the minimization of the AoI problem. The results show that our TPAoI can maintain a lower AoI while reducing the update frequency, demonstrating that our approach can keep the service status fresh at the network edge.
\end{itemize}


\section{Related Work} \label{sect-related-work}

We first classify the related work based on the phases they focus on (i.e., status updating, user accessing, and user requesting), and then discuss the efforts made to optimize AoI in dynamic environments.

\subsection{AoI Analysis of Different Phases} 
Conventional AoI was initially introduced and applied in \cite{kaul2012real} within queuing models to measure and optimize the freshness of information. These foundational queue models are mainly implemented in the updating phases to reduce transmitting delays \cite{soysal2019age, sac2018age, champati2018statistical}. 

However, conventional AoI does not consider the phases after the information is accessed, causing unnecessary updates when no user is requesting \cite{9796654}. To address this, many new metrics based on AoI have been proposed, such as EAoI (Effective AoI) \cite{eaoi2019infocom}, AoT (Age of Task) \cite{song2019age}, AoP (Age of Processing) \cite{li2021age}, QAoI (AoI at Query) \cite{qaoi2022tcomm}, and AoL (Age of Loop) \cite{de2023goal}. EAoI \cite{eaoi2019infocom} and QAoI \cite{qaoi2022tcomm} recognize that the importance of status values varies over time and is only valuable when used for decision-making, i.e., user accessing. \cite{9796654, song2019age, li2021age, chen2023joint} assert that status is freshest only when it is processed, and finally reaches the destination node. \cite{moltafet2023status, chihchun2024} on AoI optimization under bidirectional delays extend the AoI analysis to the status updating and result return phases. Meanwhile, AoL \cite{de2023goal} focuses on a closed environment and is the first to comprehensively cover various phases of status processing, introducing the concept of ``loop'' AoI to ensure the freshness of status in the sensing-control-execution loop.

To our knowledge, there has been limited research that jointly considers status updating, user accessing, and user requesting. Existing studies tend to focus on only one or two of these aspects. When only the user access process is considered, the other two phases are often overlooked, leading to status updates arriving after the user has accessed the data. Conversely, when the focus is solely on the transmitting phases, there is a tendency to send excessive and unnecessary status updates, which can increase the network burden.

\subsection{AoI Optimizing Methods in Dynamic Environment} 
Meanwhile, optimizing AoI also faces the challenge of environmental dynamics, such as delays and user behaviors. As a key enabler, reinforcement learning can model these dynamics. Related methods are classified into model-based and model-free categories based on the availability of environmental information.

For model-based methods, \cite{qaoi2022tcomm} designs a quadruple environmental state that includes AoI, data transmission error probability, node energy consumption, and user access stochastic process. It constructs precise transition probabilities based on specific distributions to optimize long-term average AoI, enhancing the freshness of the state when accessed by users. \cite{holm2023goal} studies the round-trip interaction process between sampling and control nodes, using the AoI of each round-trip stage and the round-trip link load as environmental state, and minimizes the long-term average AoI based on geometric probability distribution.

For model-free methods, \cite{holm2023goal} uses Deep Q Network (DQN) to design the mean squared error loss between observed and actual state information values as part of the reward function, maximizing long-term returns and improving the accuracy of state information. \cite{huang2023aoi} considers the sensor’s data transmission queue and the edge server’s task processing queue size as environmental state, incorporating constraints like energy consumption and service rate into the reward function. \cite{dai2022aoi}  minimizes the AoI of crowdsensing data collection by using relational graph convolutional networks (GNN) to predict environmental state transitions and Monte Carlo tree search to improve UAV path planning strategies, solving the issue of difficult-to-model state transitions. \cite{xie2021reinforcement} addresses task offloading decisions for multiple sensing nodes and multi-node task selection decisions at the server end. It extends precise MDP methods to an ANN-based model-free DQN approach, integrating the cost of multi-task selection decisions at the server into the task offloading decision cost function for multiple nodes, achieving results nearly consistent with model-based methods.

When optimizing AoI in compute-first networking, there are almost no universal solutions that involve all three phases and also address unknown dynamics. Therefore, to overcome this, we propose TPAoI and DRL-based solutions to maintain the status freshness at the network edge.

\section{Three-Phase Updating System}\label{sect-system-model}

We model the system using a time-slotted approach, with discrete time slots indexed as 1, 2, \dots, $\infty$. It is assumed that any status information arriving within a given time slot is processed at the start of the subsequent time slot. We denote the send time of the $k$th status update as $u_k$ and the corresponding arrival time at the AP as $u_k^\prime$.
As in \cite{yates2021}, we represent the conventional AoI at time slot $t$ as $\Delta(t)$, which represents the difference between the current time and the generation time of the last status update successfully received:
\begin{equation}\label{eq_aoi}
\Delta(t)=t-\max_{k:u_k^\prime\leq t}u_k.
\end{equation}

Compared to the definition above, the following equivalent definition of the conventional AoI is more useful in a time-slot system, as it delineates the variation of AoI between time slots:
\begin{equation}\label{eq_aoi_sink}
\begin{split}
\Delta(t)=&\begin{cases}\Delta(t-1)+1&\text{if }t\not=u^\prime_k\\u^\prime_k-u_k&\text{if }t=u^\prime_k\end{cases}\\
\end{split},
\end{equation}
where $\Delta(t)=0$, and $k=\underset {k} { \operatorname {arg\,max} }\, u^\prime_k\leq t$ represents the index of the status update that has most recently arrived relative to time $t$. It indicates that AoI increases linearly until a status update arrives, resetting it to $u^\prime_k-u_k$.

\begin{figure}[htbp]
\centerline{}
\includegraphics[width=0.50\textwidth]{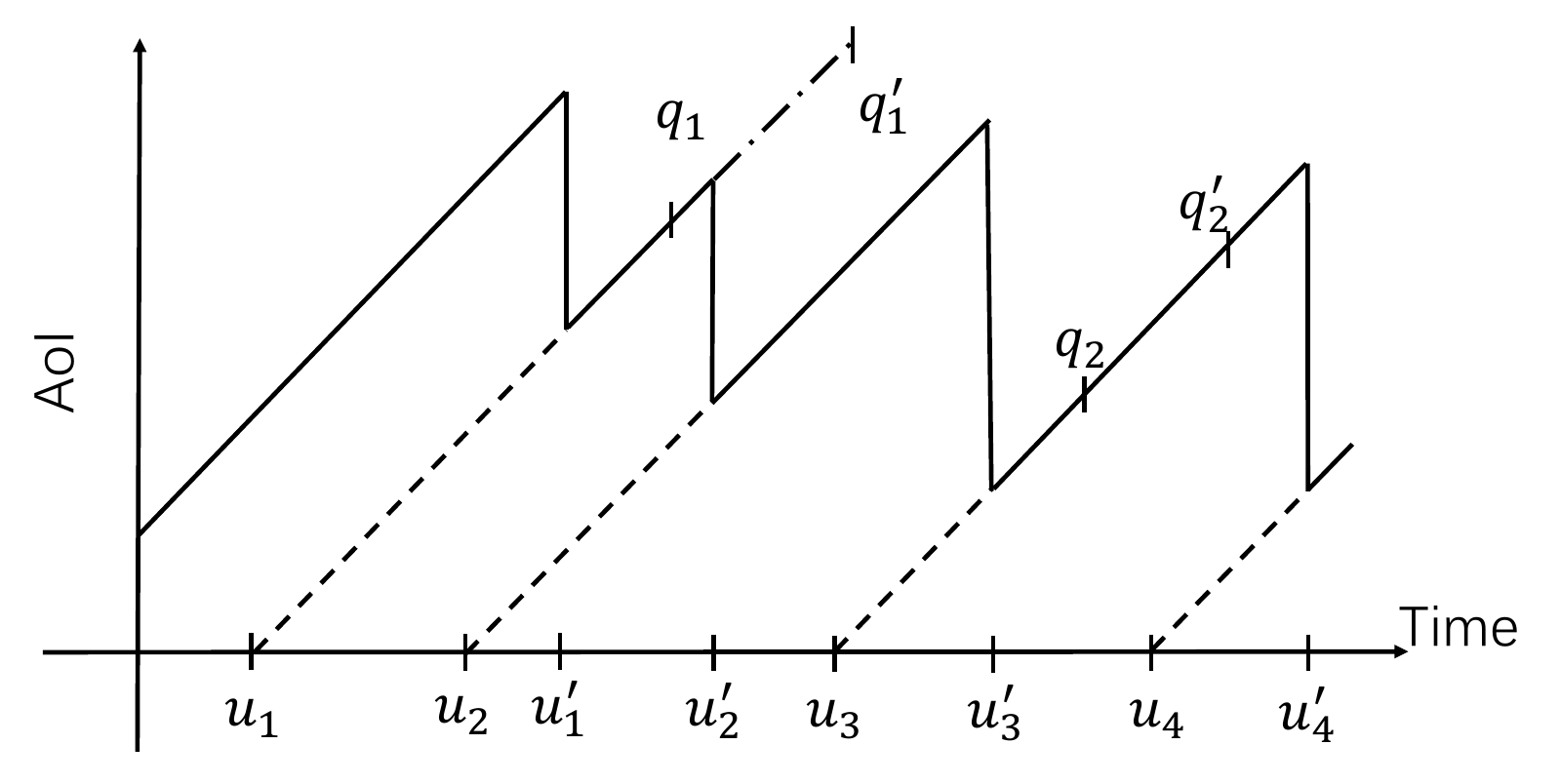}
\caption{AoI evolution in three-phase updating system}
\label{aoi}
\end{figure}

\subsection{TPAoI Metric Construction} 

We define the time when the user's $i$th request accesses the status at the AP as $q_i$. Subsequently, the request is transmitted to the remote service, arriving at time $q^\prime_i$. Then, we define \model as the AoI of the request at the time of its arrival, denoted as $\tau(i)$:
\begin{equation}\label{eq_TPAoI}
\begin{split}
\tau(i)=&\Delta(q_i)+q_i^\prime-q_i\\
=&(u^\prime_k-u_k)+(q_i-u_k^\prime)+(q_i^\prime-q_i)
\end{split},
\end{equation} 
where $k=\underset {k} { \operatorname {arg\,max} }\, u^\prime_k\leq q_i$ represents the index of the status update that has most recently arrived relative to query $q_i$. 

As in \eqref{eq_TPAoI}, \model can be divided into the sum of three components: The forwarding delay of the status update $u_k^\prime-u_k$, the waiting time for user access $q_i-\displaystyle\max_{u^\prime_k\leq q_i} u^\prime_k$, and the requesting delay $q_i^\prime-q_i$. \figurename~\ref{aoi} illustrates the process of AoI changes within the system. At $u_1$, it generates a status update; after an updating time $u_1^\prime-u_1$, the status arrives at the AP, where the AoI is then reset. Similarly, when a user accesses the status at $q_1$ and sends a request, it also requires a requesting delay $q_1^\prime-q_1$ to reach the edge server.

As can be seen in \figurename~\ref{aoi}, although the arrival time $u_2^\prime$ of the second status update reduces the conventional AoI at the AP (i.e., $\Delta(t)$), the AoI related to the first request (started from $q_1$) continues to increase until $q_1^\prime$. In other words, the second status update is unrelated to the first request and thus reducing only $\Delta(t)$ could not reduce the system’s TPAoI $\tau(i)$.

Based on \eqref{eq_TPAoI}, the long-term expected TPAoI $\tau_{\infty}$ is defined as:
\begin{equation}\label{eq_TPAoI_target}
\tau_\infty=\underset {T\to\infty} { \operatorname {lim\,sup} }\frac{1}{T}\mathbb{E}\left[\sum_{i:q_i^\prime\leq T}\tau(i)\right].
\end{equation}
Our goal is to minimize $\tau_\infty$ to decide when the server should generate and forward the status update. 

To minimize \eqref{eq_TPAoI_target}, according to \eqref{eq_TPAoI}, we need to understand those three stochastic phases to ensure that the status update arrives at the AP precisely before the user accesses it. However, due to the dynamics of the environments, knowing this information is non-trivial. We address this challenge in Sect. IV by using a DRL-based algorithm to achieve minimal TPAoI.

\subsection{Communication Model When Updating and Requesting}
 


We assume that the edge server can obtain all information in the system through telemetry or piggybacking techniques \cite{TAN2021107763}. We also simplify the network by using a first-come, first-served (FCFS) principle for transmitting status updates and user requests. 

Communication between the service and the AP includes propagation delay (at the speed of light) and transmission delay (data transmission). Due to resource competition in dynamic environments, transmission delays are subject to fluctuations, which are assumed to follow a normal\cite{1212677} or exponential\cite{4016157} distribution. 
It is worth noting that the purpose of our method is to capture various dynamics and maintain a certain degree of generality, rather than being limited to just the two distributions mentioned above.
For simplicity, we refer to both updating and requesting delays as transmission delays.

Additionally, similar to \cite{moltafet2023status}, we also adopt a wait-$N$ policy where at most $N$ status updates can be concurrently present. Considering the capacity of the system (or service), we assume there are at most $M$ requests on the link.

Due to the potential for network congestion caused by excessive update transmissions from the remote service, we impose restrictions on their transmission frequency to save costs. Specifically, we introduce a penalty mechanism, where sending an update incurs a negative reward. This encourages the server to avoid transmitting updates unless necessary. As illustrated in \figurename~\ref{aoi}, because there is no user request, the transmission of update $u_2$ is deemed undesirable and can be omitted to conserve network bandwidth.

\section{TPAoI Minimization Based on DRL}

In this section, we formalize the system model outlined in Sect. \ref{sect-system-model} using a MDP. This approach enables us to effectively adapt to the dynamic environments. An MDP is characterized by the tuple $(\mathcal{S}, \mathcal{A}, \mathcal{P}, \mathcal{R})$, where $\mathcal{S}$ represents the state space, indicating the state of the system; $\mathcal{A}$ represents the action space, denoting the actions the server can take; $\mathcal{P}$ denotes the state transition probabilities $P(\boldsymbol{s_{t+1}}|a_t,\boldsymbol{s_t})$; and $\mathcal{R}$ represents the reward function $R(\boldsymbol{s_{t}}, a_{t}, \boldsymbol{s_{t+1}})$, indicating the immediate reward received after an action and the subsequent state transition.

\subsection{State $\mathcal{S}$ Representation for Three Phases}
\begin{figure}[ht]
    \centering
    \includegraphics[width=0.49\textwidth]{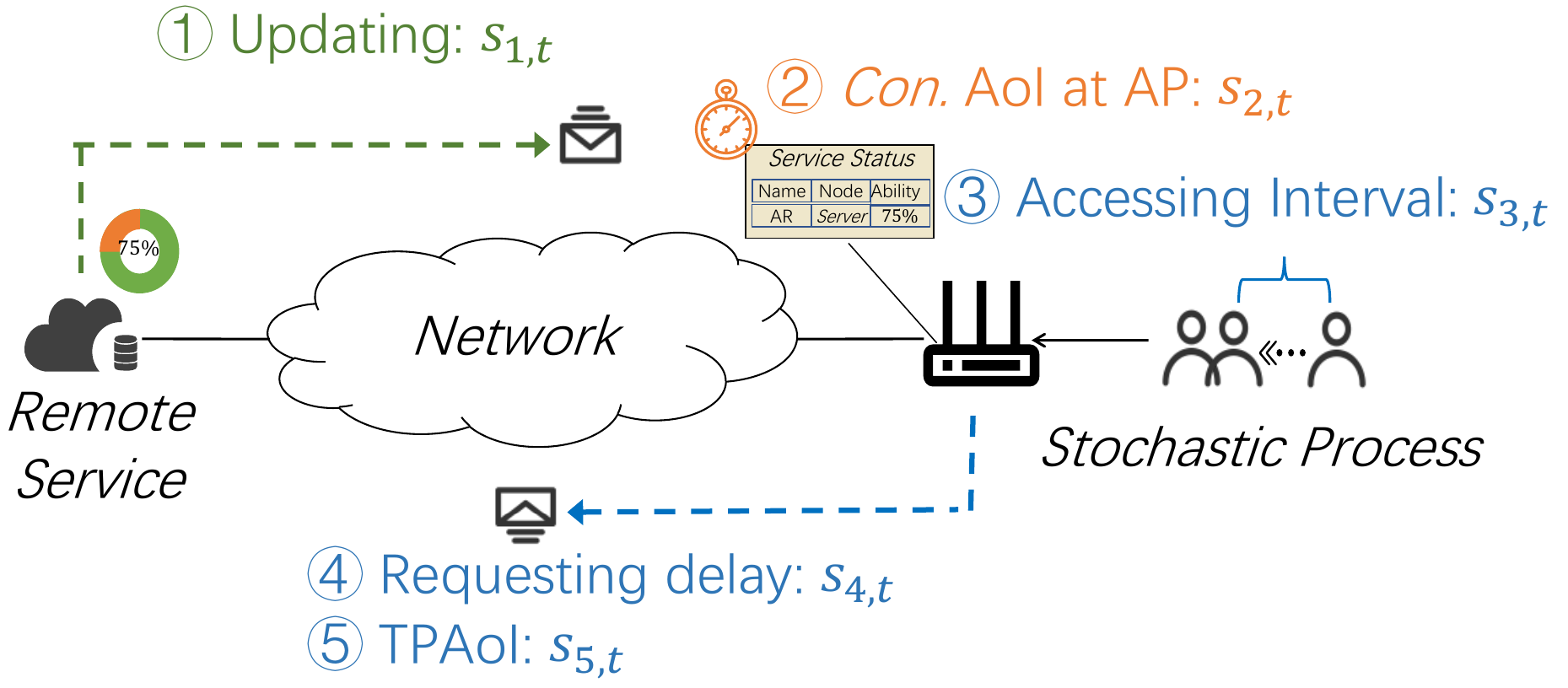} 
    \caption{Illustration of state $\mathcal{S}$ \textit{w.r.t} three phases}
    \label{state}
\end{figure}

To fully capture the dynamics of the entire process, we have designed five sub-states. Three of these sub-states capture the three stochastic phases, while the remaining two are related to the AoI, specifically the conventional AoI and the TPAoI. The five sub-states are illustrated as follows:

\begin{enumerate}[label=\protect\circled{\arabic*}]
    \item $\boldsymbol{s_{1,t}}$ represents the transmission time for status updates, illustrating the stochastic process of the first phase (i.e., the status update phase), and it is an N-dimensional vector.
    \item $s_{2,t}$ represents the conventional AoI (\textit{Con.} AoI for short) at the AP, 
    which is the waiting period for user access.
    \item $s_{3,t}$ represents the interval between user accesses, illustrating the stochastic process of the second phase, which is the user access phase.
    \item $\boldsymbol{s_{4,t}}$ represents the transmission time of user requests, illustrating the stochastic process of the third phase (i.e., the user request phase), and it is an M-dimensional vector.
    \item $\boldsymbol{s_{5,t}}$ represents the AoI of the request, and it is an M-dimensional vector. Upon the arrival of the request, this value becomes TPAoI.
\end{enumerate}%
These five sub-states together form a comprehensive state $\boldsymbol{s_t}$ that effectively captures the dynamic changes of TPAoI in the system.
\figurename~\ref{state} illustrates the corresponding positions of each state. We discuss them in detail as follows.

Firstly, we need to introduce the AoI of the AP, denoted as $s_{2,t}$, so that we can update it when users are about to access and the AoI is high. The definition of $s_{2,t}$ is the same as the definition of conventional AoI, which is:
\begin{equation}
    s_{2,t}=\Delta(t).
\end{equation}
However, due to transmission delays, whether a status update arrives at the current time is related to whether edge servers sent status updates in the previous time slots. In other words, the next state is related to actions taken several time slots ago, which does not align with the properties of MDP, where the next state should depend only on the current state and not on previous states:
\begin{equation}\label{eq_mdp}
P(\boldsymbol{s_{t+1}}|a_{t},\boldsymbol{s_t},\ldots,a_1,\boldsymbol{s_1})=P(s_{t+1}|a_{t},\boldsymbol{s_t}).
\end{equation}


Secondly, to ensure the state aligns with MDP properties, we introduce a state for each status update on the link to represent its transmission time, thereby determining when the AP updates its AoI. These states of update collectively form $\boldsymbol{s_{1,t}}$. 
Additionally, we allow a maximum of $N$ status updates concurrently on the link to determine the length of $\boldsymbol{s_{1,t}}$ and avoid training difficulties due to the variable length of $\boldsymbol{s_{1,t}}$. Here is the definition of the $i$th ($i \leq N$) state value in $\boldsymbol{s_{1,t}}$: 
\begin{equation}
s_{1,t+1,i}=\begin{cases}
    s_{1,t,i}+1&\text{if }s_{1,t,i}\not=0\wedge z_{t,i}=0\\
    1&\text{if }a_t=1\wedge x=i\\
    0&\text{otherwise}
\end{cases},
\end{equation}
where $z_{t, i}\in\{0,1\}$ denote whether $i$th status update has arrived at the AP at time slot $t$ with 1 indicating that it has arrived, and $x$ is a random sample drawn from the set that includes each index $j$ meets $s_{1,t,j} = 0$, $j=1,\dots, N$. Here is the explanation of the formula: A value of 0 indicates an idle state. When an edge server sends a status update, it randomly selects a state with a value of 0 and sets it to 1. Otherwise, the state represents the transmission time of the status update on the link. It increments by 1 each time the status update hasn't arrived and resets to 0 upon arrival. 

Thirdly, to estimate the time user access, we introduce a state to represent the interval between user accesses, denoted as $s_{3,t}$, which linearly increases and resets to 1 upon user arrival. The $s_{3,t}$ can be expressed as follows:
\begin{equation}
s_{3,t+1}=\begin{cases}
    s_{3,t}+1&\text{if }y_t=0\\
    1&\text{if }y_t=1
\end{cases},
\end{equation}
where $y_t\in\{0,1\}$ denote whether the user accesses at time slot $t$. 

Fourthly and finally, we introduced two states for each request on the link to represent its transmission time and its AoI, respectively, to ensure that the reward function conforms to the properties of the MDP:
\begin{equation}
P(r_{t}|a_{t},\boldsymbol{s_t},\ldots,a_1,\boldsymbol{s_1})=P(r_{t}|a_{t},\boldsymbol{s_t}).
\end{equation}
The sets of transmission time and AoI respectively form $\boldsymbol{s_{4,t}}$ and $\boldsymbol{s_{5,t}}$. Here is the definition of the $i$th ($i\leq M$) state value in $\boldsymbol{s_{4,t}}$ and $\boldsymbol{s_{5,t}}$:
\begin{equation}
s_{4,t+1,i}=\begin{cases}
    s_{4,t,i}+1&\text{if }s_{4,t,i}\not=0\wedge k_{t,i}=0\\
    1&\text{if }y_t=1\wedge x=i\\
    0&\text{otherwise}
\end{cases},
\end{equation}
\begin{equation}
s_{5,t+1,i}=\begin{cases}
    s_{5,t,i}+1&\text{if }s_{5,t,i}\not=0\wedge k_{t,i}=0\\
    s_{2,t}+1&\text{if }y_t=1\wedge x=i\\
    0&\text{otherwise}
\end{cases},
\end{equation}
where $k_{t, i}\in\{0,1\}$ denote whether the $i$th request has arrived at the edge server at time slot $t$, and $x$ is a random sample drawn from the set that includes each index $j$ meets $s_{1,t,j} = 0$, $j=1,\dots, M$. The definition of the request’s transmission time value is the same as that of the status update. The AoI of the request is 0 when no request is sent, set to the current AoI upon sending a request, increased by 1 in each time slot, and reset to 0 upon arrival. To keep the state size fixed, we limit the maximum number of requests on the link to $M$. However, $M$ should be set relatively large to prevent the user from occupying all states simultaneously.

According to our illustration above, the state space of $\mathcal{S}$ is $\mathbb{N}^{N+2+M\times 2}$. 

\subsection{Updating Action $\mathcal{A}$ and Reward $\mathcal{R}$}

The edge server can decide whether to send a status update in each time slot. Accordingly, the action space is defined as $\mathcal{A}=\{0,1\}$, where $a_t$ denotes the action taken at time slot $t$. Specifically, $a_t = 1$ signifies that the edge server opts to send a status update during that time slot, while $a_t = 0$ indicates that no status update is sent.

The reward is defined as the AoI of the request upon its arrival at the edge server. Additionally, to prevent the edge server from sending status updates ineffectively, we assign a negative reward $\omega$ when it sends a status update. The reward at time slot $t$, denoted as $r_t$, can be represented as follows: 
\begin{equation}\label{eq_reward}
r_t=-\big(\omega\cdot a_t+\sum_{i=0}^M r_{t, i}\big),
\end{equation}
\begin{equation}
    r_{t,i}=s_{5,t,i}\cdot k_{t,i}.
\end{equation}

After modeling the MDP, our overall optimization objective \eqref{eq_TPAoI_target} becomes finding a policy $\pi^*$ to maximize the expected state value:
\begin{equation}\label{eq_target}
    \pi^*=\underset {\pi} { \operatorname {arg\,max} }\ \mathbb{E}[V_\pi(\boldsymbol{s})|\boldsymbol{s_0}],
\end{equation}
where $V_\pi(s)$ is the state value that denotes the expected discounted cumulative reward, i.e.:
\begin{equation}\label{eq_state_value}
    V_\pi(s)=\mathbb{E}[G_t|\boldsymbol{s_t}=\boldsymbol{s},\pi],
\end{equation}
\begin{equation}\label{eq_return}
    G_t=\sum^\infty_{\tau=0}(\gamma)^\tau r_{t+\tau+1},
\end{equation}
where $\gamma < 1$ is the discount factor, which ensures $V_\pi$ convergence. Additionally, the value of $\gamma$ can control the range of the rewards considered. Due to the dynamics existing, transition probabilities implied in the value of state are usually difficult to obtain in practical applications. We use a DRL-based algorithm to fit these probabilities by continuously interacting with the environment.

\subsection{DRL-Based TPAoI Minimization Algorithm}

In this section, we thoroughly explore how to apply a DRL-based algorithm, namely D3QN, to solve the MDP defined earlier. Specifically, the Q-value function employs a Dueling Network architecture\cite{wang2016dueling}, while loss computation and training are implemented using Double DQN\cite{van2016deep}.

\begin{algorithm}[H]
\caption{DRL-Based Optimal Algorithm}
\label{alg}
\begin{algorithmic}
\State Initialize $\boldsymbol{s_1}, \epsilon, \epsilon_{decay}, \epsilon_{min}, \gamma, \eta, T_0$;
\State Initialize parameter of Q-network $\boldsymbol{\theta}$,
\State Initialize parameter of target Q-network $\boldsymbol{\theta}^-$
\State Initialize RB
\For{$t = 1, 2, \dots$}
    \State Input $\boldsymbol{s_t}$ to obtain $q(\boldsymbol{s_t}, a)$ accorting to \eqref{eq_q_learning}.
    \State Acquire $a_t$ accorting to \eqref{eq_optimal}.
    \State Execute action $a_t$ according to $\epsilon$-greedy policy
    \State Observe $\boldsymbol{s_{t+1}}, r_{t}$
    \State Store $(\boldsymbol{s_t}, a_t, \boldsymbol{s_{t+1}}, r_{t})$ into RB.
    \If{Remainder $(t, T_0) == 0$}
        \State Randomly sample a batch experiences $\Xi$ from RB
    \For{each $\boldsymbol{\xi} = (\boldsymbol{s_t}, a_t, \boldsymbol{s_{t+1}}, r_{t}) \in \Xi$}
        \State Calculate $y_t$ by \eqref{eq_target_value}.
    \EndFor
    \State Calculate Loss by \eqref{eq_loss}.
    \State Update $\boldsymbol{\theta}$ according to \eqref{eq_sgd}.
    \State Update the target Q-network according to \eqref{soft_update}.
    \State Update the $\epsilon$ value according to \eqref{eq_decy}.
    \EndIf
\EndFor
\end{algorithmic}
\end{algorithm}

Unlike DQN\cite{mnih2015human}, we use two deep neural networks to approximate the Q-value function, typically employing plain deep neural networks for construction:
\begin{equation}\label{eq_q_learning}
\begin{split}
Q_{\boldsymbol{\theta}}(\boldsymbol{s},a)=&\mathbb{E}[G_t|\boldsymbol{s_t}=\boldsymbol{s},a_t=a,\pi^*],\\
=&V_{\boldsymbol{\theta}_1}(\boldsymbol{s})+A_{\boldsymbol{\theta}_2}(\boldsymbol{s},a)-\underset {a} { \operatorname {mean} }\,A_{\boldsymbol{\theta}_2}(\boldsymbol{s},a)
\end{split},
\end{equation}
where $\boldsymbol{\theta}$ represents the parameters of the neural network, $V_{\boldsymbol{\theta}_1} (\cdot)$ approximates the state value function, and $A_{\boldsymbol{\theta}_2}(\cdot)$ approximates the advantage function. Compared to DQN, which directly approximates the Q-value function, this algorithm incorporates a more refined structure that enables faster convergence. 

Since the Q-value function represents the expected discounted cumulative reward following action, the optimal action can be determined by achieving the maximum Q-value in the state:
\begin{equation}\label{eq_optimal}
    a^*(\boldsymbol{s})=\underset{a\in\mathcal{A}}{\operatorname{arg\,max}}\,Q_{\boldsymbol{\theta}}(\boldsymbol{s}, a).
\end{equation}

Just like DQN, Double DQN also requires the use of two networks: The Q-network and the target Q-network, denoted respectively as $Q_{\boldsymbol{\theta}}(\boldsymbol{s}, a)$ and $Q_{\boldsymbol{\theta}^-}(\boldsymbol{s}, a)$, with $\boldsymbol{\theta}$ and $\boldsymbol{\theta}^-$ as their corresponding parameters. Initially, the target Q-network and the Q-network have the same structure and parameters. The target Q-network is employed to compute the Temporal Difference (TD) target, enabling stable training of the network. Additionally, compared to DQN, Double DQN addresses the problem of overestimation by using two separate estimations to calculate the TD error during training.

During the training process, we leverage the experience replay technique to maximize the utilization of training data. Specifically, we use a Replay Buffer (RB) to store historical experiences as 4-tuples, denoted by $\boldsymbol{\xi}=(\boldsymbol{s_t},a_t,\boldsymbol{s_{t+1}},r_{t})$. We let the RB as a circular queue with a maximum length of $L$. Training is conducted once every $T_0$ iteration. During training sessions, a mini-batch of experiences $\boldsymbol{\Xi}$ is sampled from the RB to update the parameters $\boldsymbol{\theta}$ by minimizing the following loss, namely the expected square of TD error:
\begin{equation}\label{eq_loss}
loss=\mathbb{E}_{\boldsymbol{\xi}_t\sim\boldsymbol{\Xi}}\bigg[\big(y_t-Q_{\boldsymbol{\theta}}(\boldsymbol{s_t},a_t)\big)^2\bigg],
\end{equation}
where $y_t$ denotes the TD target:
\begin{equation}\label{eq_target_value}
y_t=r_t+\gamma Q_{\boldsymbol{\theta}^-}\left(\boldsymbol{s_{t+1}},\arg\max_{a_t}Q_{\boldsymbol{\theta}}\left(\boldsymbol{s_{t+1}},a_t\right)\right).
\end{equation}

We use stochastic gradient descent (SGD)\cite{bottou2012stochastic} to minimize the loss, which simplifies the process of computing the gradient of the expectation over the entire dataset by using a single training sample or a mini-batch of samples in each iteration:
\begin{equation}\label{eq_sgd}
\boldsymbol{\theta}\leftarrow\boldsymbol{\theta}-\eta\bigg[\big(Q_{\boldsymbol{\theta}}(\boldsymbol{s_t},a_t)-y_t\big)\nabla_{\boldsymbol{\theta}}Q_{\boldsymbol{\theta}}(\boldsymbol{s_t},a_t)\bigg],
\end{equation}
where $\eta$ is the learning rate. Furthermore, we employ a soft update technique to gradually adjust the parameters $\boldsymbol{\theta}^-$:
\begin{equation}\label{soft_update}
\boldsymbol{\theta}^-=\tau\boldsymbol{\theta} + (1-\tau)\boldsymbol{\theta}^-,
\end{equation}
where $\tau$ controls the rate at which the weights of the target network are updated. To enhance exploration, we employ the $\epsilon$-greedy strategy, where actions are chosen at random from the action space with a probability of $\epsilon$, and \eqref{eq_optimal} is followed with a probability of $1-\epsilon$, where $\epsilon$ decays according to the following formula, thereby increasing the exploitation:
\begin{equation}\label{eq_decy}
\epsilon=\max(\epsilon\cdot\epsilon_{decay}, \epsilon_{min}),
\end{equation}
where $\epsilon_{decay}$ represents the decay factor, and $\epsilon_{min}$ represents the minimum value of $\epsilon$. Finally, we summarize the DRL-
based optimal in Algorithm \ref{alg}.

\section{Experiments and Analysis}

In this section, we demonstrate the effectiveness of our proposed method in minimizing TPAoI. Specifically, we consider the following four aspects: 
\begin{enumerate}[label=\protect\circled{\arabic*}]
    \item We compare TPAoI with QAoI and conventional AoI to illustrate the superiority of TPAoI.
    \item We explain how TPAoI updates before user access by considering transmission delays, thereby achieving the effect of minimizing AoI at the time of user access.
    \item We conduct ablation studies on the transmission cost and the number of concurrently existing status updates to illustrate their impact on the results.
    \item We present the training curve of our model to verify its convergence.
\end{enumerate}

\subsection{Experiment Settings}

\begin{table}[htbp]
\caption{Parameters for experiments}
\centering
\begin{tabular}{c|c}
\hline
\textbf{Parameter} & \textbf{Value}\\
\hline
accessing process & 20+$Poisson(\lambda$)\\
\hline
transmission delay fluctuation&  $exp(1)$ or $\mathcal{N}(1,1)$ \\
\hline
transmission cost $\omega$&  1\\
\hline
discounted factor $\gamma$&  0.995\\
\hline
learning rate $\eta$& 0.0002 \\
\hline
replay buffer length $L$&  50,000\\
\hline
greedy factor $\epsilon$& 1 \\
\hline
$\epsilon_{decay}$&  0.98\\
\hline
$\epsilon_{min}$& 0.01 \\
 \hline
 hidden layers&  [128,512,256,128]\\
\hline
\end{tabular}
\label{tab1}
\end{table}

\noindent\textbf{(1) Network Settings.}
We consider the following edge computing network scenario (see \figurename~\ref{state} for network topology): 
In the user accessing phase, according to \cite{qaoi2022tcomm} we set the user access interval at 20 time slots. Moreover, to create a dynamic environment, we introduce a Poisson distribution with the user accessing dynamic parameter $\lambda$ on this basis. A lower $\lambda$ means user access frequency is lower, which means it's hard to predict the user's behavior, and the system dynamics are increased. 
Additionally, To simulate the network dynamics, transmitting delay fluctuations are modeled using either an exponential distribution with a mean of 1 or a normal distribution with both mean and variance of 1, denoted as $exp(1)$ and $\mathcal{N}(1,1)$, respectively. 

\noindent\textbf{(2) DRL Settings.}
The value function and advantage function in the Q-network have similar network structures but differ in their output dimensions, which are 1 and the dimension of the action space, respectively. 
The network includes four hidden layers with neuron counts of 128, 512, 256, and 128, utilizing the ReLU activation function. The learning rate $\eta$ is set to 0.0002, with a discount factor $\gamma$ of 0.995. The capacity of the RB $L$ is 50,000, and each mini-batch $\Xi$ contains 128 samples. The soft update parameter $\tau$ is set at 0.001. The training encompasses 5000 episodes, with a transmission cost $\omega$ of 1 per transmission. The initial greedy factor $\epsilon$ is 1, which decays at a rate of 0.98 per episode until it reaches 0.01. The maximum number of status updates present on the link at any given time is capped at two. Additionally, considering that the interval between user accesses significantly exceeds the transmission delay of requests, it is sufficient to allow only one request to exist on the link concurrently. \tablename~\ref{tab1} shows the parameters used in the experiments.


\begin{figure*}[ht]
\centering
\subfigure[Transmission delay fluctuations following $exp(1)$]{\label{exp_compare}
\includegraphics[width=0.49\textwidth]{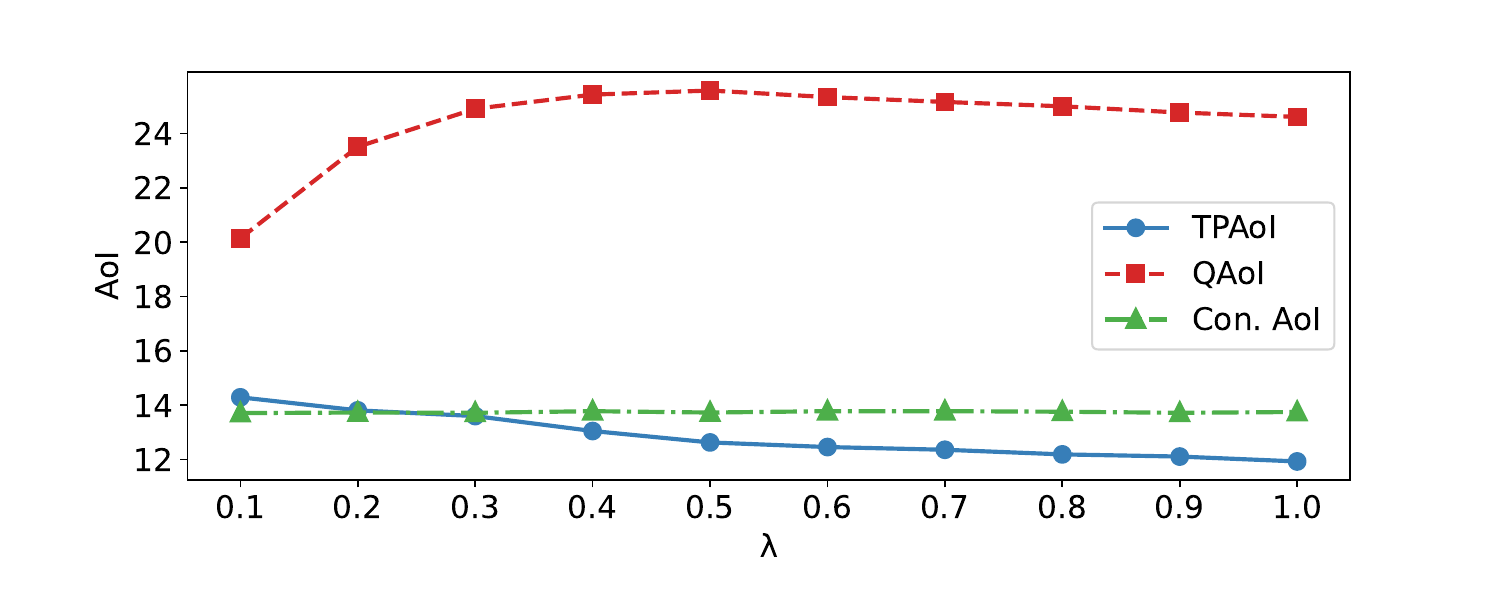}}
\subfigure[Transmission delay fluctuations following $\mathcal{N}(1,1)$]{\label{normal_compare}
\includegraphics[width=0.49\textwidth]{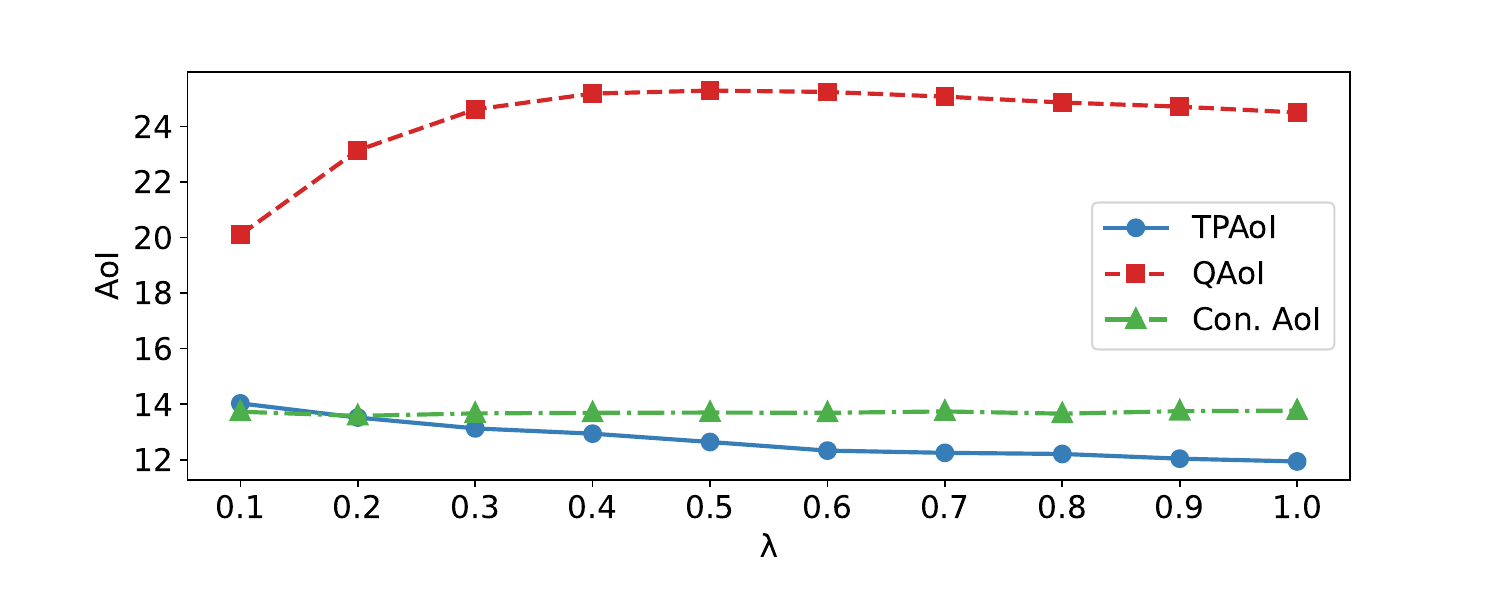}}

\subfigure[Corresponding total number of status updates of Fig. \ref{exp_compare}]{\label{energy_exp}
\includegraphics[width=0.49\textwidth]{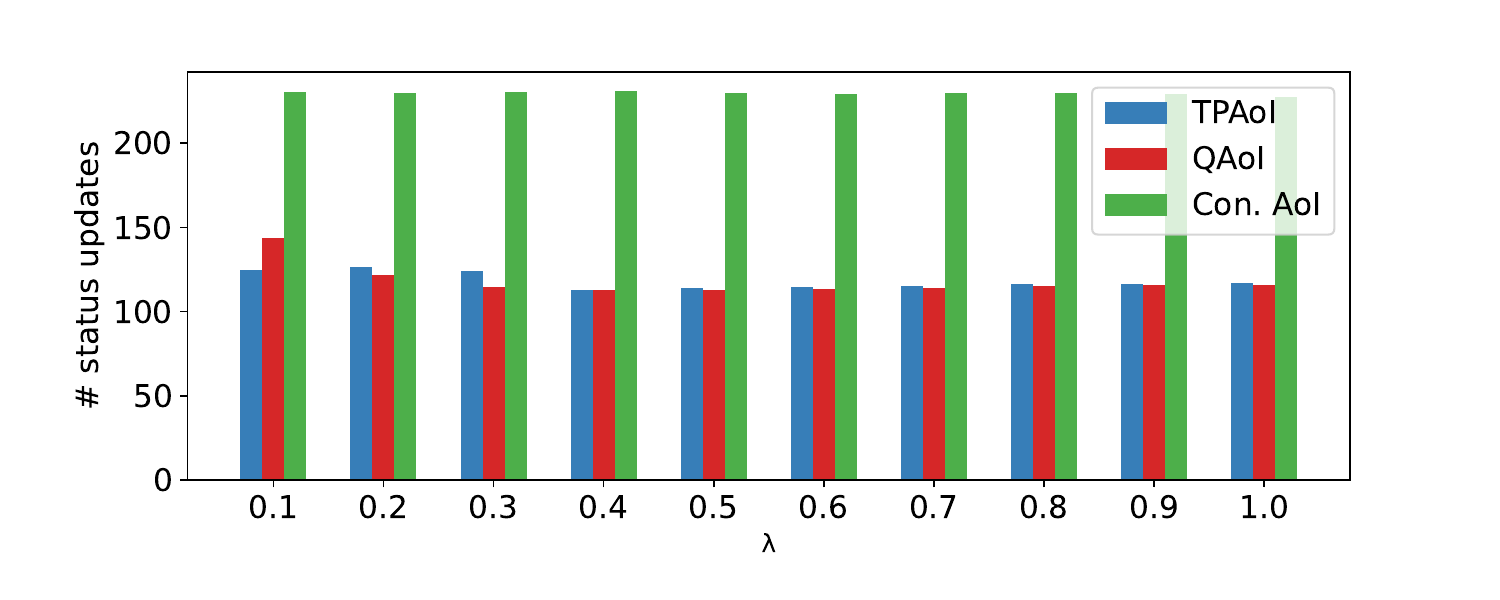}}
\subfigure[Corresponding total number of status updates of Fig. \ref{normal_compare}]{\label{energy_normal}
\includegraphics[width=0.49\textwidth]{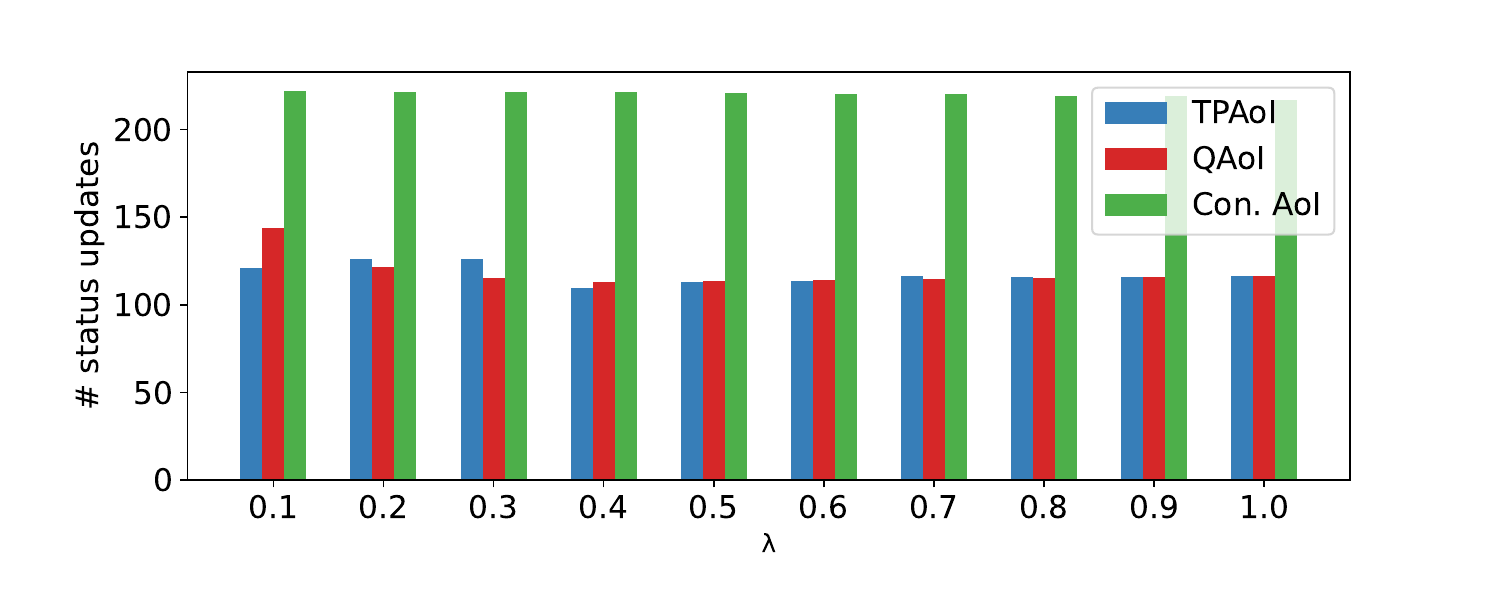}}
\caption{Performance comparison of TPAoI, QAoI, and Con. AoI varying $\lambda$. (a) and (b) show the AoI comparison when transmitting delay fluctuation following $exp(1)$ and $\mathcal{N}(1,1)$, respectively. (c) and (d) show the corresponding total number of transmitted status updates from (a) and (b), respectively.}
\label{baseline_compare}
\end{figure*} 

\noindent\textbf{(3) Baselines.}
We choose the most recent QAoI \cite{qaoi2022tcomm} and the conventional AoI \cite{yates2021} as our baselines.
\begin{itemize}
    \item QAoI: This model focuses solely on the AoI at the time of user access and can capture the user dynamics, disregarding forwarding and requesting delays. It assumes that status updates sent in the current time slot arrive at the access point immediately in the next time slot. QAoI is fundamentally a simplified version of the TPAoI model, having removed the transmission delay and only retaining the state of the user access interval and AoI. Due to its simplicity, we use the policy iteration method for training.
    \item Conventional AoI (Con. AoI for short): It assumes that users may access information at any given moment, which does not accurately simulate actual user behavior. AoI can be considered a TPAoI model with a fixed user access interval of 1; therefore, we omit the user access interval in the state definition and utilize the same optimization algorithm used for TPAoI.
\end{itemize}

We then deploy the trained QAoI and AoI models in the TPAoI environment to compare and analyze the performance.

\subsection{Performance Evaluation}

In this section, we demonstrate how our metric comprehensively considers both AoI and the total number of status updates (i.e., the number of transmissions), as shown in \figurename~\ref{baseline_compare}. 

We first illustrate \figurename~\ref{exp_compare} and \ref{energy_exp}, where the transmission delay fluctuation follows an exponential distribution and user access follows a Poisson distribution with different $\lambda$. 
As shown in \figurename~\ref{exp_compare}, in most cases TPAoI gets a lower AoI. Compared with QAoI, the AoI of TPAoI is significantly lower, improving about 47\% on average. Putting \figurename~\ref{energy_exp} altogether, which shows the corresponding total number of status updates of \figurename~\ref{exp_compare}, we find that TPAoI can not only ensure the status fresh but also keeps the number of status updates at a low level. In the comparison between TPAoI and conventional AoI, TPAoI has far fewer status updates on average about 48\% than conventional AoI, demonstrating that TPAoI sends status updates only when necessary, thereby achieving higher efficiency.  

As the parameter $\lambda$  gradually increases, the randomness of user access continuously decreases. This change directly leads to a monotonic decrease in the AoI for TPAoI. When $\lambda$ is low, the main reason for the higher AoI is the limitation of transmission costs, which prevents the model from adapting to this increased randomness by increasing the frequency of status updates. Additionally, the higher AoI for QAoI is primarily because QAoI focuses on the arrival process of users but overlooks the impact of transmission delays, resulting in status updates being completed only after the user arrives. Conventional AoI's performance remains relatively stable because it measures the AoI at any moment; thus, as long as the user access process is periodic, the results remain unchanged.  

\figurename~\ref{normal_compare} and \ref{energy_normal} present the comparison of AoI and the number of status updates respectively when transmission delay fluctuation follows a normal distribution. Similarly to \figurename~\ref{exp_compare} and \ref{energy_exp}, TPAoI obtains better results, indicating that our model is capable of adapting to different transmission delay fluctuations and user dynamic behaviors.

In summary, the results indicate that our method not only optimizes the AoI, achieving superior timeliness of information, but also effectively controls the frequency of status updates, avoiding excessive communication. 

Given the similar results between the exponential and normal distributions, and considering space limitations, subsequent experiments default to displaying results for the exponential distribution unless otherwise specified.

\subsection{Correctness Evaluation}


In this section, we compare TPAoI with adjusted QAoI and analyze the distribution graph of AoI relative to user access intervals in TPAoI, to illustrate how effectively TPAoI fits the dynamics that existed in the accessing phase.

\begin{figure}[htbp]
    \centering
    \includegraphics[width=0.49\textwidth]{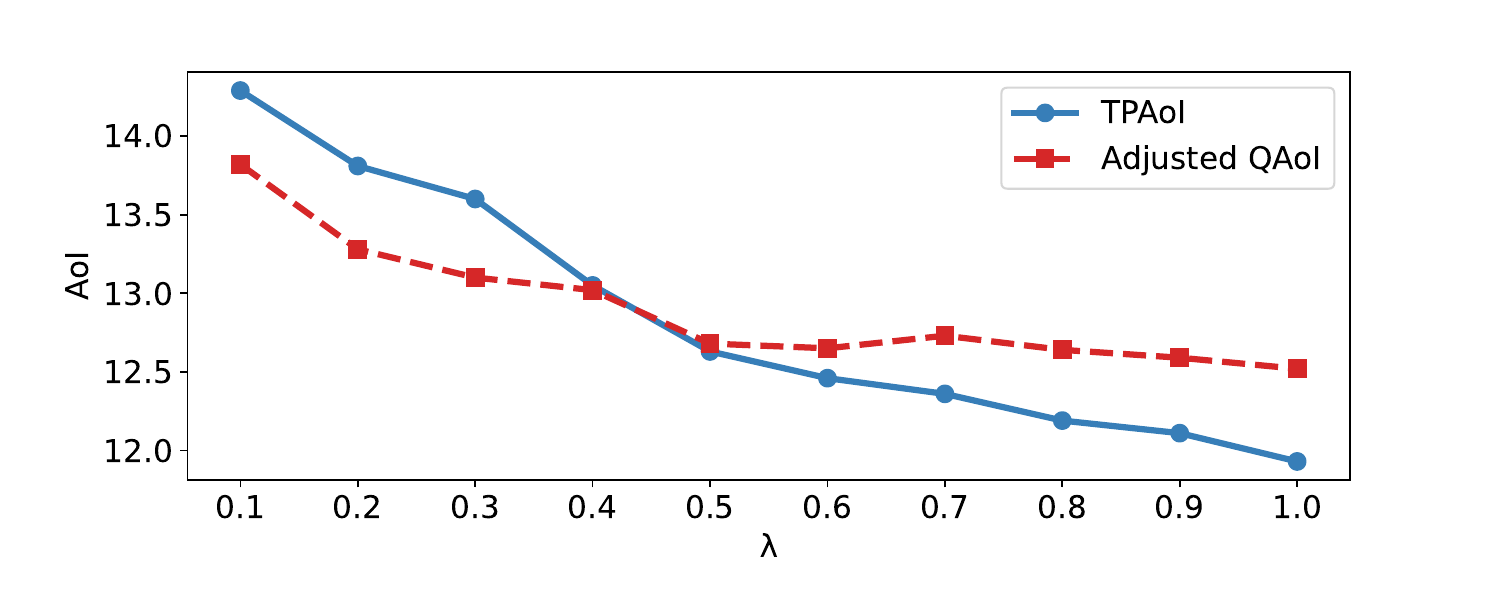} 
    \caption{AoI  fitting $w.r.t$ $\lambda$}
    \label{offset_compare}
\end{figure}

In our experiment settings, according to the network topology, the total expected transmission time is slightly less than 12. Considering that fitting the probability distribution of network fluctuation may lead to some losses, we set the expected transmission time to 12. To align QAoI's transmission time with this expectation, we adjust it by adding an offset value, referring to it as the \textit{adjusted QAoI}. To some extent, we use the adjusted QAoI as the correct reference.

\figurename~\ref{offset_compare} provides a detailed comparison between TPAoI and the adjusted QAoI. We can see that both TPAoI and the adjusted QAoI metrics exhibit a decreasing trend as $\lambda$ increases. On average, the difference in AoI between TPAoI and the adjusted QAoI is only approximately 0.366. This indicates that, in terms of AoI, our algorithm effectively models the user access process, thereby optimizing AoI performance. 
It can be observed that when $\lambda$ is low, the adjusted QAoI has a lower AoI than the TPAoI. This is because, at low $\lambda$ values, adjusted QAoI sends more status updates, whereas TPAoI maintains a relatively constant number of status updates due to the system model's limitation on the number of concurrent updates on the link.



To verify whether TPAoI correctly guides the server's updating decisions (i.e., sending the status to AP before user access), we further count the updating frequency distribution concerning the accessing interval. Specifically, we conducted multiple simulations under conditions with $\lambda=1.0$ and $\lambda=0.1$ to collect tuples (user access interval $s_3$, AoI $s_2$), and to record their frequencies to describe the distribution.
We scale the frequencies by dividing them by the maximum occurrence of access interval, to normalize them between 0 and 1. 

\begin{figure}[htbp]
\centering
\subfigure[$\lambda=1.0$]{\label{user_access_1.0}
\includegraphics[width=0.49\textwidth]{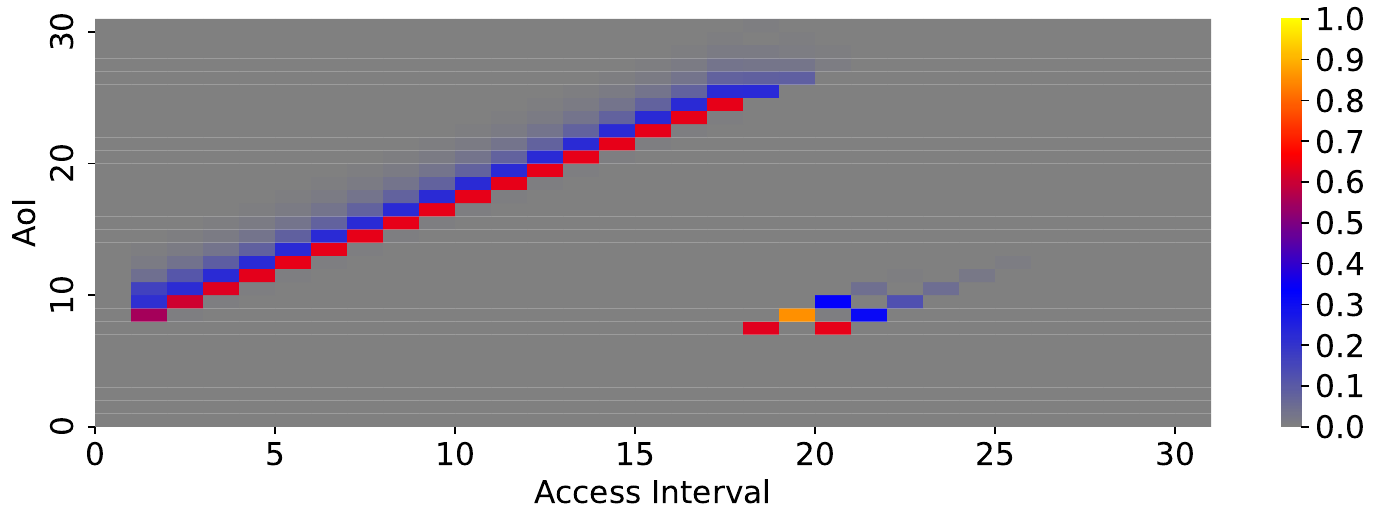}}
\subfigure[$\lambda=0.1$]{\label{user_access_0.1}
\includegraphics[width=0.49\textwidth]{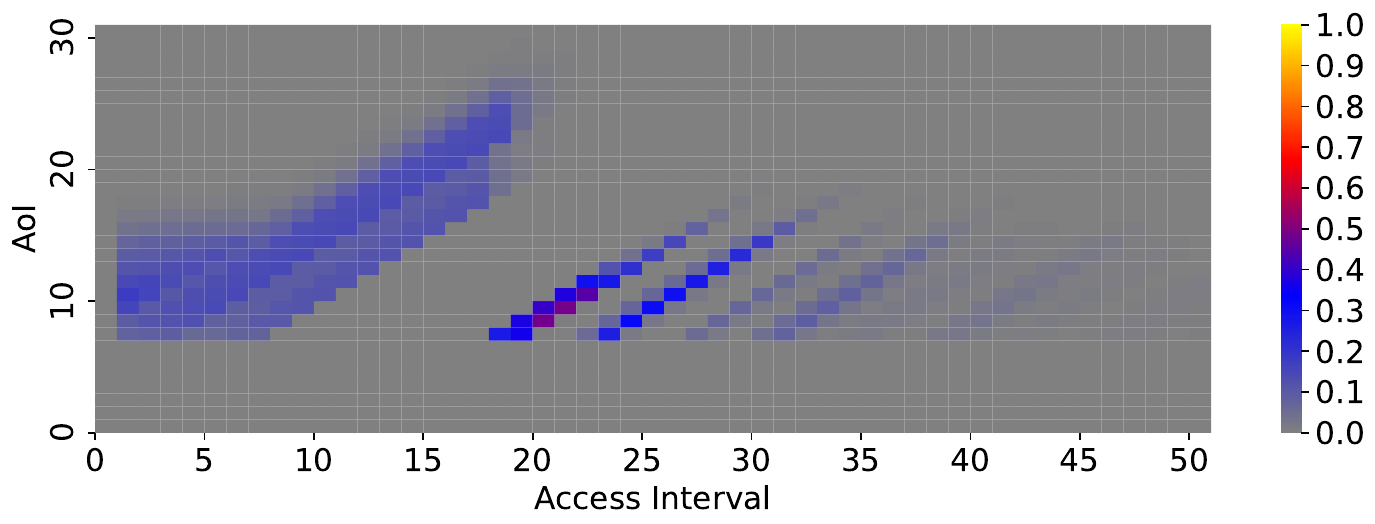}}
\caption{Frequency distribution of AoI relative to user access intervals when $\lambda=1.0$ and $\lambda=0.1$}
\label{user_access}
\end{figure}

The distribution of these tuples is displayed in \figurename~\ref{user_access}, where different colors represent different frequencies.
Taking \figurename~\ref{user_access_1.0} as an example, we analyzed the AoI evolution. Typically, users access the system when AoI is between 7 and 9, after which the access interval is reset to 1. Therefore, in most cases, when the access interval is 1, AoI is mostly between 8 and 10. As the interval increases from 1 to 15, AoI also increases, indicating that no status updates have been received during this period. When the interval increases from 15 to 20, the distribution of AoI shows two distinct parts, corresponding to situations with and without status updates. Notably, when the interval is at 20, AoI is usually distributed between 7 and 9, which marks the beginning of random access by users. This demonstrates that our model can complete status updates before a user accesses, thereby minimizing AoI at the time of user access.

\figurename~\ref{user_access_0.1} illustrates the scenario when the randomness of user access is greater, i.e., $\lambda = 0.1$. It indicates that the status update arrives at AP before the user starts random access, i.e., at an interval of 20, and continuously updates before user access, thereby minimizing the AoI. Due to the increased randomness of user access, the AoI distribution becomes broader at intervals of 1 to 15, which is because the range of AoI during user access also becomes wider.

Overall, \figurename~\ref{offset_compare} compares QAoI and AoI to highlight differences from the perspective of AoI, while \figurename~\ref{user_access} analyzes the specific changes in AoI and the timing of status update arrivals. These analyses demonstrate that our algorithm can adapt to the user's accessing phases and update the AoI before their access, demonstrating that our model can handle different levels of randomness with strong fitting capabilities.
The ability to handle different transmission delay fluctuations and random user access patterns further implies that our proposed algorithm can adapt to dynamic network conditions. Ultimately, our approach ensures that the status of the AP remains fresh at a low cost.

\subsection{Ablation}

\begin{figure}[ht]
\centering
\subfigure[Different transmission costs]{\label{send_consume}
\includegraphics[width=0.48\textwidth]{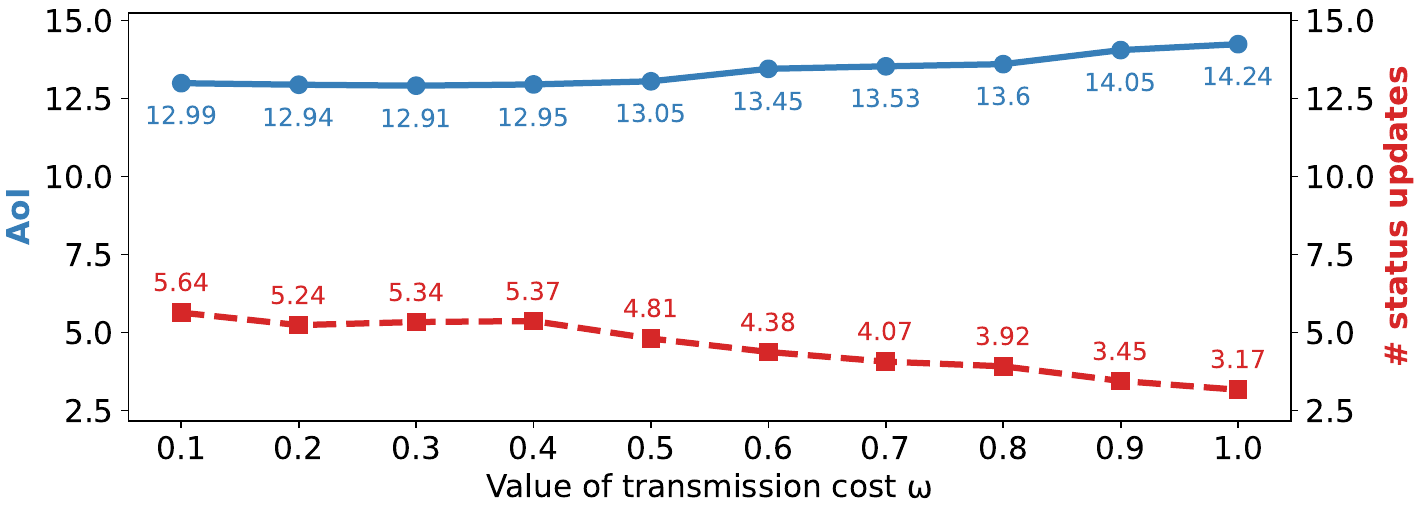}}
\subfigure[Different number of concurrent status update]{\label{update_status}
\includegraphics[width=0.48\textwidth]{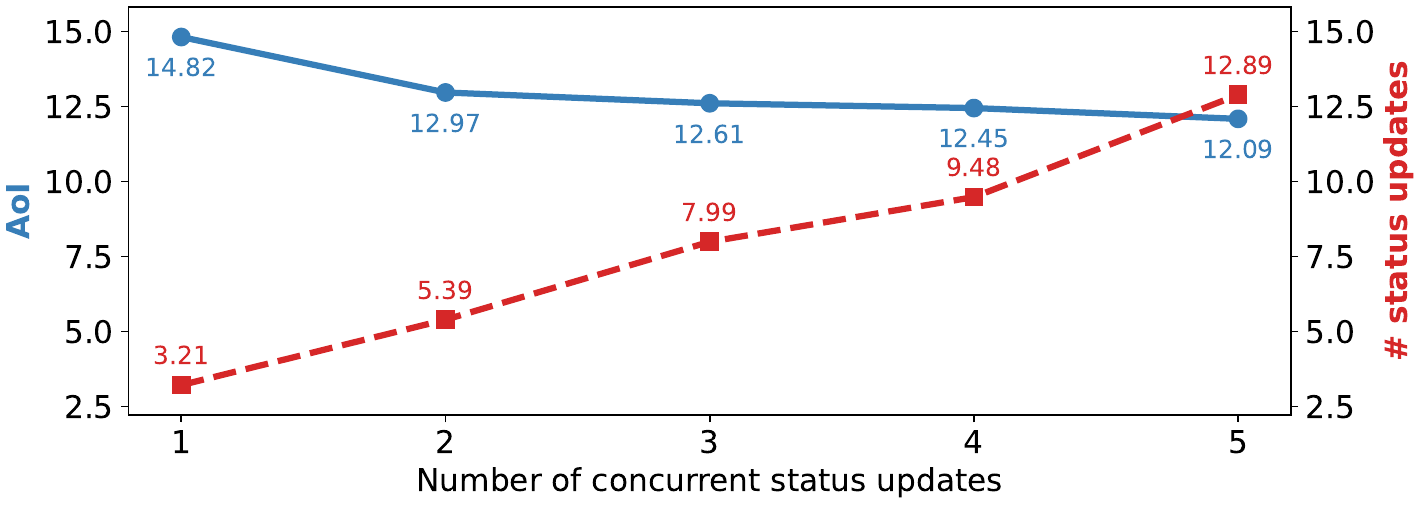}}
\caption{AoI and the number of status updates under different transmission costs and the number of concurrent status updates, where the number of status updates is expressed as the total number of transmissions divided by the number of user visits.}
\label{ablation}
\end{figure}

In this section, we detail the impacts of different transmission costs and the number of concurrent status updates on our model. Through \figurename~\ref{send_consume}, we can see that as the transmission cost increases, the AoI curve also rises, while the number of status updates correspondingly decreases. This indicates that when the transmission cost is low, the model tends to send more updates to reduce AoI. In other words, our model can dynamically adjust the update frequency based on different transmission costs. It can be observed that AoI is inversely proportional to the number of status updates. This is reasonable because, theoretically, as the number of status updates increases, AoI should decrease accordingly. However, it is worth noting that when the transmission cost is less than or equal to 0.4, increasing the frequency of status updates does not significantly affect AoI fluctuation. The reason for this phenomenon is that in our system setup, the number of status updates present on the link concurrently is limited to two or fewer, which directly restricts the frequency of AoI updates.

To delve deeper into the specific impact of the number of concurrent status updates on the AoI, we intentionally set the transmission cost to 0.1 to allow the model to update status more freely. This setting enables us to more effectively test the impact of different number of status updates on system performance. The results, displayed in \figurename~\ref{update_status}, clearly show that as the number of status updates increases, the AoI significantly decreases without the slight fluctuation observed in \figurename~\ref{send_consume}. This finding explicitly indicates that reducing the restrictions on the number of status updates enables our model to choose to send more updates, thereby effectively lowering the AoI.

In summary, reducing transmission costs allows our model to update status more freely, while increasing the number of concurrent status updates on the link can increase the frequency of AoI updates, both of which can reduce the AoI.

\subsection{Convergence Explanation}

\begin{figure}[htbp]
\centerline{}
\includegraphics[width=0.49\textwidth]{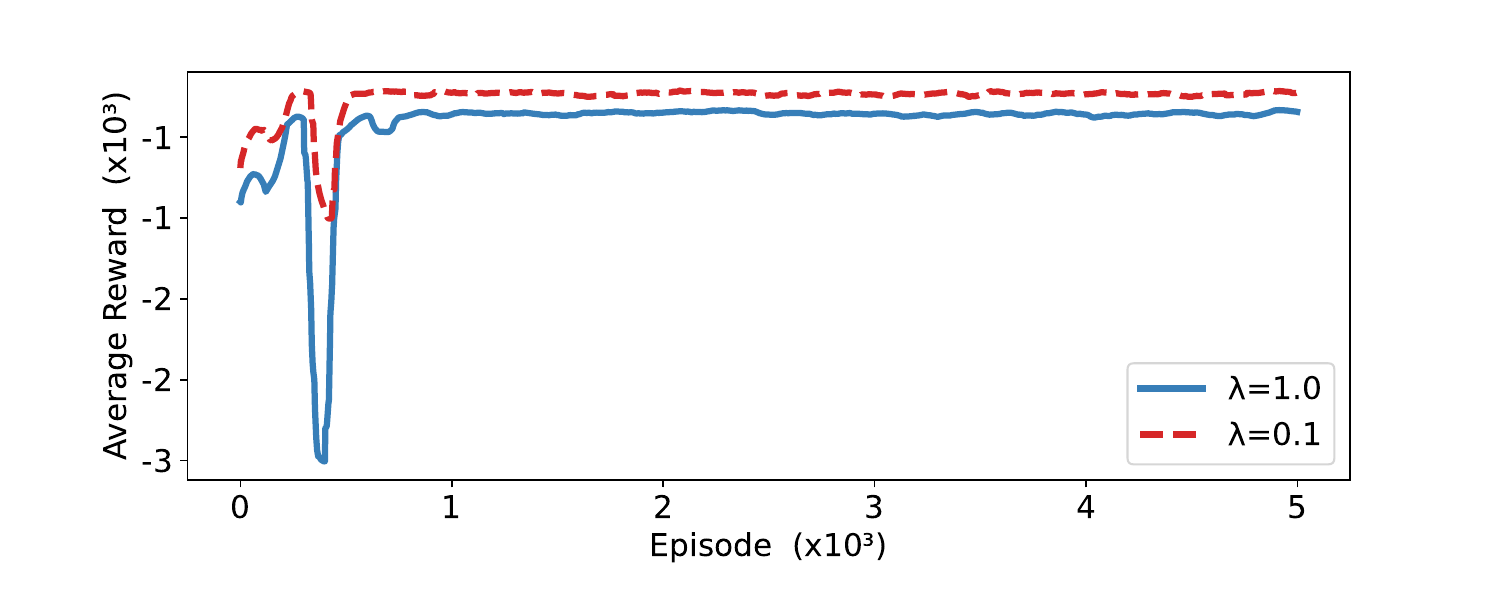}
\caption{Learning curve with parameters $\lambda=0.1$ and $\lambda=1.0$}
\label{convergence}
\end{figure}

In this section, we illustrate the convergent nature of our model by showcasing the learning curves. To clarify this further, we present the moving average results, which are obtained by averaging the rewards over consecutive 100 episodes. Specifically, as observed from \figurename~\ref{convergence}, after about 1000 episodes, the average reward tends to stabilize and gradually approaches a horizontal line. This phenomenon indicates that our algorithm exhibits good convergence properties.

Meanwhile, in the initial stages of training, we can observe significant fluctuation in rewards, primarily due to the influence of the $\epsilon$-greedy strategy. Since the $\epsilon$-greedy strategy favors exploring a variety of action choices early in training, it often opts to send updates to gain a broader range of experiences. Therefore, during the initial stages with a higher $\epsilon$ value, the model engages in necessary exploratory activities, leading to some fluctuation. Furthermore, as the $\epsilon$ value starts to decrease, the model also explores the consequences of not sending updates, which results in a decrease in the average reward. Over time, as the model accumulates sufficient experience and gradually adjusts the $\epsilon$ value, its performance becomes more stable and begins to demonstrate the effectiveness of the optimization process.

\section{Conclusion}

In this paper, we introduce a new metric called TPAoI, which comprehensively considers the stochastic distribution of status update delay, user access behavior, and request delay. This holistic measurement fully captures the timeliness and accuracy of information, particularly in dynamic network environments. We formulate the TPAoI minimization problem as a MDP and propose a DRL-based optimization algorithm. Extensive experimental validation demonstrates that our model effectively optimizes TPAoI while reducing update frequency. Our research contributes a novel metric to enhance status-updating systems in the compute-first networking domain, ensuring effective status management at a relatively lower operational cost. Future work will focus on extending our solution to encompass multiple edge servers and access points.

\section*{Acknowledgment}
The authors would like to thank the anonymous reviewers for their valuable comments and suggestions, which helped to improve the quality of this paper.
\balance
\bibliographystyle{plain}
\bibliography{references}

\begin{thebibliography}{10}

\bibitem{abbas2023comprehensive}
Qamar Abbas, Syed~Ali Hassan, Hassaan~Khaliq Qureshi, Kapal Dev, and Haejoon Jung.
\newblock A comprehensive survey on age of information in massive iot networks.
\newblock {\em Computer Communications}, 197:199--213, 2023.

\bibitem{bottou2012stochastic}
L{\'e}on Bottou.
\newblock Stochastic gradient descent tricks.
\newblock In {\em Neural Networks: Tricks of the Trade: Second Edition}, pages 421--436. Springer, 2012.

\bibitem{champati2018statistical}
Jaya~Prakash Champati, Hussein Al-Zubaidy, and James Gross.
\newblock Statistical guarantee optimization for age of information for the {D/G/1} queue.
\newblock In {\em IEEE INFOCOM 2018-IEEE Conference on Computer Communications Workshops (INFOCOM WKSHPS)}, pages 130--135. IEEE, 2018.

\bibitem{chen2023joint}
Yi~Chen, Zheng Chang, Geyong Min, Shiwen Mao, and Timo H{\"a}m{\"a}l{\"a}inen.
\newblock Joint optimization of sensing and computation for status update in mobile edge computing systems.
\newblock {\em IEEE Transactions on Wireless Communications}, 22(11):8230--8243, 2023.

\bibitem{qaoi2022tcomm}
Federico Chiariotti, Josefine Holm, Anders~E. Kalør, Beatriz Soret, Søren~K. Jensen, Torben~B. Pedersen, and Petar Popovski.
\newblock Query age of information: Freshness in pull-based communication.
\newblock {\em IEEE Transactions on Communications}, 70(3):1606--1622, 2022.

\bibitem{dai2022aoi}
Zipeng Dai, Chi~Harold Liu, Yuxiao Ye, Rui Han, Ye~Yuan, Guoren Wang, and Jian Tang.
\newblock {AoI-minimal UAV crowdsensing by model-based graph convolutional reinforcement learning}.
\newblock In {\em IEEE INFOCOM 2022-IEEE Conference on Computer Communications}, pages 1029--1038. IEEE, 2022.

\bibitem{de2023goal}
Pedro~M de~Sant~Ana, Nikolaj Marchenko, Beatriz Soret, and Petar Popovski.
\newblock Goal-oriented wireless communication for a remotely controlled autonomous guided vehicle.
\newblock {\em IEEE Wireless Communications Letters}, 12(4):605--609, 2023.

\bibitem{9796654}
Xingqiu He, Sheng Wang, Xiong Wang, Shizhong Xu, and Jing Ren.
\newblock Age-based scheduling for monitoring and control applications in mobile edge computing systems.
\newblock In {\em IEEE INFOCOM 2022 - IEEE Conference on Computer Communications}, pages 1009--1018, 2022.

\bibitem{holm2023goal}
Josefine Holm, Federico Chiariotti, Anders~E Kal{\o}r, Beatriz Soret, Torben~Bach Pedersen, and Petar Popovski.
\newblock Goal-oriented scheduling in sensor networks with application timing awareness.
\newblock {\em IEEE Transactions on Communications}, 71(8):4513--4527, 2023.

\bibitem{huang2023aoi}
Jiwei Huang, Han Gao, Shaohua Wan, and Ying Chen.
\newblock {AoI-aware energy control and computation offloading for industrial IoT}.
\newblock {\em Future Generation Computer Systems}, 139:29--37, 2023.

\bibitem{kaul2012real}
Sanjit Kaul, Roy Yates, and Marco Gruteser.
\newblock Real-time status: How often should one update?
\newblock In {\em 2012 Proceedings IEEE INFOCOM}, pages 2731--2735. IEEE, 2012.

\bibitem{cfn2019icn}
Micha\l{} Kr\'{o}l, Spyridon Mastorakis, David Oran, and Dirk Kutscher.
\newblock Compute first networking: Distributed computing meets {ICN}.
\newblock In {\em Proceedings of the 6th ACM Conference on Information-Centric Networking}, ICN '19, page 67–77, New York, NY, USA, 2019. ACM.

\bibitem{li2021age}
Rui Li, Qian Ma, Jie Gong, Zhi Zhou, and Xu~Chen.
\newblock {Age of processing: Age-driven status sampling and processing offloading for edge-computing-enabled real-time IoT applications}.
\newblock {\em IEEE Internet of Things Journal}, 8(19):14471--14484, 2021.

\bibitem{ling2022age}
Zhuang Ling, Fengye Hu, Hongliang Zhang, and Zhu Han.
\newblock Age-of-information minimization in healthcare {IoT} using distributionally robust optimization.
\newblock {\em IEEE Internet of Things Journal}, 9(17):16154--16167, 2022.

\bibitem{mnih2015human}
Volodymyr Mnih, Koray Kavukcuoglu, David Silver, Andrei~A Rusu, Joel Veness, Marc~G Bellemare, Alex Graves, Martin Riedmiller, Andreas~K Fidjeland, Georg Ostrovski, et~al.
\newblock Human-level control through deep reinforcement learning.
\newblock {\em Nature}, 518(7540):529--533, 2015.

\bibitem{moltafet2023status}
Mohammad Moltafet, Markus Leinonen, Marian Codreanu, and Roy~D Yates.
\newblock Status update control and analysis under two-way delay.
\newblock {\em IEEE/ACM Transactions on Networking}, 31(6):2918--2933, 2023.

\bibitem{jianpeng2024}
Jianpeng Qi, Xiao Su, and Rui Wang.
\newblock Towards distributively build time-sensitive-service coverage in compute first networking.
\newblock {\em IEEE/ACM Transactions on Networking}, 32(1):582--597, 2024.

\bibitem{sac2018age}
Hakan Sac, Tan Bacinoglu, Elif Uysal-Biyikoglu, and Giuseppe Durisi.
\newblock Age-optimal channel coding blocklength for an {M/G/1} queue with {HARQ}.
\newblock In {\em 2018 IEEE 19th International Workshop on Signal Processing Advances in Wireless Communications (SPAWC)}, pages 1--5. IEEE, 2018.

\bibitem{1212677}
Meng-Fu Shih and A.O. Hero.
\newblock Unicast-based inference of network link delay distributions with finite mixture models.
\newblock {\em IEEE Transactions on Signal Processing}, 51(8):2219--2228, 2003.

\bibitem{song2019age}
Xianxin Song, Xiaoqi Qin, Yunzheng Tao, Baoling Liu, and Ping Zhang.
\newblock Age based task scheduling and computation offloading in mobile-edge computing systems.
\newblock In {\em 2019 IEEE Wireless Communications and Networking Conference Workshop (WCNCW)}, pages 1--6. IEEE, 2019.

\bibitem{soysal2019age}
Alkan Soysal and Sennur Ulukus.
\newblock Age of information in {G/G/1/1} systems.
\newblock In {\em 2019 53rd Asilomar Conference on Signals, Systems, and Computers}, pages 2022--2027. IEEE, 2019.

\bibitem{easiei2024iotj}
Xiao Su, Jianpeng Qi, Jiahao Wang, and Rui Wang.
\newblock Easiei: A simulator to flexibly modeling complex edge computing environments.
\newblock {\em IEEE Internet of Things Journal}, 11(1):1558--1571, June 2024.

\bibitem{TAN2021107763}
Lizhuang Tan, Wei Su, Wei Zhang, Jianhui Lv, Zhenyi Zhang, Jingying Miao, Xiaoxi Liu, and Na~Li.
\newblock In-band network telemetry: A survey.
\newblock {\em Computer Networks}, 186:107763, 2021.

\bibitem{van2016deep}
Hado Van~Hasselt, Arthur Guez, and David Silver.
\newblock {Deep reinforcement learning with double Q-learning}.
\newblock In {\em Proceedings of the AAAI conference on artificial intelligence}, volume~30, 2016.

\bibitem{chihchun2024}
C.~C. Wang.
\newblock Optimal aoi for systems with queueing delay in both forward and backward directions.
\newblock {\em IEEE/ACM Transactions on Networking}, pages 1--16, 2024.

\bibitem{wang2016dueling}
Ziyu Wang, Tom Schaul, Matteo Hessel, Hado Hasselt, Marc Lanctot, and Nando Freitas.
\newblock Dueling network architectures for deep reinforcement learning.
\newblock In {\em International Conference on Machine Learning}, pages 1995--2003. PMLR, 2016.

\bibitem{4016157}
Ye~Xia and David Tse.
\newblock Inference of link delay in communication networks.
\newblock {\em IEEE Journal on Selected Areas in Communications}, 24(12):2235--2248, 2006.

\bibitem{xie2021reinforcement}
Xin Xie, Heng Wang, and Mingjiang Weng.
\newblock A reinforcement learning approach for optimizing the age-of-computing-enabled {IoT}.
\newblock {\em IEEE Internet of Things Journal}, 9(4):2778--2786, 2021.

\bibitem{xu2022aoi}
Chengyuan Xu, Qian Xu, Jianping Wang, Kui Wu, Kejie Lu, and Chunming Qiao.
\newblock {AoI-centric task scheduling for autonomous driving systems}.
\newblock In {\em IEEE INFOCOM 2022-IEEE Conference on Computer Communications}, pages 1019--1028. IEEE, 2022.

\bibitem{yates2021}
Roy~D. Yates, Yin Sun, D.~Richard Brown, Sanjit~K. Kaul, Eytan Modiano, and Sennur Ulukus.
\newblock Age of information: An introduction and survey.
\newblock {\em IEEE Transactions on Communications}, 39(5):1183--1210, 2021.

\bibitem{eaoi2019infocom}
Bo~Yin, Shuai Zhang, Yu~Cheng, Lin~X. Cai, Zhiyuan Jiang, Sheng Zhou, and Zhisheng Niu.
\newblock Only those requested count: Proactive scheduling policies for minimizing effective age-of-information.
\newblock In {\em IEEE INFOCOM 2019 - IEEE Conference on Computer Communications}, pages 109--117, 2019.

\bibitem{cpnsurvey2024}
Sun Yukun, Lei Bo, Liu Juniin, Huang Haonan, Zhang Xing, Peng Jing, and Wang Wenbo.
\newblock Computing power network: A survey.
\newblock {\em China Communications}, pages 1--37, 2024.

\end{thebibliography}

\end{document}